\documentstyle[aps,amssymb,epsfig]{revtex}
\begin{document}
\title
{ Topological Confinement of Spins and Charges:\\ 
Spinons as ${\large\pi-}$ junctions.}
\author{\large{ S. Brazovskii }$^{1,2}$}
\bigskip
\address{
$^1$ Laboratoire de Physique Th\'eorique et de s Mod\`eles Statistiques, CNRS;\\
 B\^at.100, Universit\'e Paris-Sud, 91405 Orsay cedex, France;\\
 http://ipnweb.in2p3.fr/~lptms/membres/brazov/, 
e-mail:  brazov@ipno.in2p3.fr }
\address{$^2$ L.D. Landau Institute, Moscow, Russia.}
\maketitle
\begin{abstract}  
 Topologically nontrivial states, the solitons, emerge as elementary excitations 
in $1D$ electronic systems. In a quasi $1D$ material the topological 
requirements originate the spin- or charge- roton like excitations with charge- 
or spin- kinks localized in the core. They result from the spin-charge 
recombination due to confinement and the combined symmetry. The rotons possess 
semi-integer winding numbers which may be relevant to configurations discussed 
in connection to quantum computing schemes. Practically important is the case of 
the spinon functioning as the single electronic $\pi -$ junction in a quasi $1D$ 
superconducting material. (Published in \cite{moriond}.)
\end{abstract}

\section{Introduction to solitons.}

Topological defects: solitons, vortices, anyons, etc., are discussed currently, 
see  \cite{ivanov}, in connection to new trends in physics of quantum devices, 
see \cite{esteve}. Closest to applications and particularly addressed at this 
conference \cite{ryazanov} are the $\pi -$ junctions which, linking  two 
superconductors,  provide degeneracy of their states with phase differences 
equal to  $0$ and  $2\pi$. The final goal of this publication is to show that in 
quasi one-dimensional ($1D$) superconductors the $\pi -$ junctions are produced 
already at the single electronic level extendible to a finite spin polarization. 
The effect results from reconciliation of the spin and the charge which have 
been separated at the single chain level. The charge and the spin of the single 
electron reconfine as soon as $2D$ or $3D$ long range correlations are 
established due to interchain coupling.  The phenomenon is much more general 
taking place, in other respects, also in such a  common system as the Charge 
Density Wave -CDW and in such a popular system as the doped antiferromagnet or 
the Spin Density Wave as its quasi $1D$ version \cite{braz-00}. Actually in this 
article we shall consider firstly and in greater details the CDW which is an 
object a bit distant to the mesoscopic community. The applications to 
superconductors will become apparent afterwards.
We shall concentrate on effects of interchain coupling $D>1$: confinement, 
topological constraints, combined symmetry, spin-charge recombination. A short 
review and basic references on history of solitons and related topics in 
correlated electronic systems (like holes moving within the antiferromagnetic media)
can be found in \cite{braz-00}. 

Solitons in superconducting wires were considered very early \cite{AL}, within 
the macroscopic regime of the Ginzburg - Landau theory, for the phase slips 
problem. Closer to our goals is the microscopic solution for solitonic lattice 
in quasi $1D$ superconductors \cite{buzdin} at the Zeeman magnetic field.  
This successful application of results from theory of CDWs, see \cite{SB-84}, 
to superconductors  provides  also a link of pair breaking effects 
in these different systems. The solitonic 
structures in  qasi $1D$ superconductors appear as a $1D$ version of the well 
known FFLO (Fulde, Ferrel, Larkin, Ovchinnikov, see \cite{buzdin}) 
inhomogeneous state near the pair breaking limit. Being very weak in $3D$, this effect becomes quite 
pronounced in systems with nested Fermi surfaces which is the case of the $1D$ 
limit. 

To extend physics of solitons to the higher $D$ world, the most important 
problem one faces is the effect of \emph{confinement} (S.B. 1980): 
as topological objects connecting degenerate vacuums, the solitons at $D>1$ 
acquire an infinite energy unless they reduce or compensate their topological 
charges. The problem is generic to solitons but it becomes particularly 
interesting at the electronic level where the spin-charge reconfinement 
appears as the result of topological constraints.
The topological effects of $D>1$ ordering reconfines the charge and the spin 
\emph{locally} while still with essentially different distributions. 
Nevertheless \emph{integrally} one of the two is screened again, being 
transferred to the collective mode, so that in transport the massive particles 
carry only either charge or spin as in $1D$, see reviews \cite{SB-84,SB-89}.

\section{Confinement and combined excitations.}

\subsection{The classical commensurate CDW: confinement of phase solitons and of 
kinks.}

The CDWs were always considered as the most natural electronic systems to 
observe solitons. We shall devote to them some more attention because the CDWs also 
became the subject of studies in mesoscopics \cite{delft}. 
Being a case of spontaneous symmetry breaking, the CDW order parameter 
$O_{cdw}\sim\Delta \cos [{\bf{Qr}}+\varphi ]$ possesses a manifold of degenerate
ground states. For the $M-$ fold commensurate CDW the energy 
$\sim \cos [M\varphi]$ reduces the allowed positions to multiples of 
$2\pi /M$, $M>1$.
Connecting trajectories $\varphi \rightarrow \varphi \pm 2\pi /M$ are phase 
solitons, or ''$\varphi -$ particles'' after Bishop \textit{et al}. Particularly 
important is the case $M=2$ for which solitons are clearly observed e.g. in 
polyacethylene \cite{PhToday} or in organic Mott insulators \cite{monceau}. 

Above the $3D$ or $2D$ transition temperature $T_{c}$, the symmetry is not 
broken and solitons are allowed to exist as elementary particles. But in the 
symmetry broken phase at $T<T_{c}$, any local deformation must return the 
configuration to the same (modulo $2\pi $ for the phase) state. Otherwise the interchain 
interaction energy (with the linear density 
$F\sim \left\langle\Delta _{0}\Delta _{n}\cos [\varphi _{0}-
\varphi _{n}]\right\rangle $ 
is lost when the effective phase 
$\varphi _{0}+\pi sign(\Delta _{0})$
at the soliton bearing chain $n=0$ acquires a finite (and $\neq 2\pi )$ 
increment with respect to the neighboring chain values $\varphi _{n}$. The $1D$ 
allowed solitons do not satisfy this condition which originates a constant 
\textit{confinement force} $F$ between them, hence the infinitely growing 
confinement energy $F|x|$. E.g. for $M=2$ the kinks should be bound in pairs or 
aggregate into macroscopic complexes with a particular role plaid by Coulomb 
interactions \cite{teber}.

Especially interesting is the more complicated case of {\it coexisting 
discrete and continuous symmetries}. As a result of their interference the 
topolological charge of solitons originated by the discrete symmetry can be  
compensated by gapless degrees of freedom originated by the continuous one. This 
scenario we shall discuss through the rest of the article.

\subsection{ The incommensurate CDW:\\
 confinement of Amplitude Solitons with phase wings.}

Difference of ground states with even and odd numbers of particles is a common 
issue in mesoscopics. In CDWs it also shows up in a spectacular way
(S.B. 1980, see \cite{SB-84,SB-89}. 
Thus any pair of electrons or holes is accommodated to the extended ground states for 
which the overall phase difference becomes $\pm 2\pi$. Phase increments are produced 
by phase slips which provide the spectral flow \cite{yak} from the upper 
$+\Delta_0$ to the lower $-\Delta_0$ rims of the single particle gap. The phase 
slip requires for the amplitude $\Delta(x,t)$ to pass through zero, at which moment 
the complex order parameter has a shape of the amplitude soliton (AS, the kink
$\Delta (x=-\infty)\leftrightarrow -\Delta (x=\infty$). 
Curiously, this instantaneous configuration becomes the stationary ground state 
for the case when only one electron is added to the system or when the total 
spin polarization is controlled to be nonzero, see Figure 1.   
The AS carries the singly occupied mid-gap state, thus having a spin $1/2$ but 
its charge is compensated to zero  by local dilatation of singlet vacuum states 
\cite{SB-84,SB-89}. 

As a nontrivial topological object ($O_{cdw}$ does not map onto itself) the pure 
AS is prohibited in $D>1$ environment. Nevertheless it becomes allowed even 
their if it acquires phase tails with the total increment $\delta\varphi =\pi $, 
see Figure 2. The length of these tails $\xi _{\varphi }$ is determined by the 
weak interchain coupling, thus $\xi _{\varphi }\gg \xi _{0}$. As in $1D$, the 
sign of $\Delta$ changes within the scale $\xi _{0}$ but further on, at the 
scale $\xi _{\varphi }$, the factor $\cos [Qx+\varphi ]$ also changes
the sign thus leaving the product in $O_{cdw}$ to be invariant. As a result
the 3D allowed particle is formed with the AS core $\xi _{0}$ carrying the spin 
and the two phase $\pi /2$ twisting wings stretched over $\xi _{\varphi }$, each 
carrying the charge $e/2$. 

\begin{figure}[tbp]
\begin{minipage}[c]{.45\linewidth}
\includegraphics*[width=5.5cm]{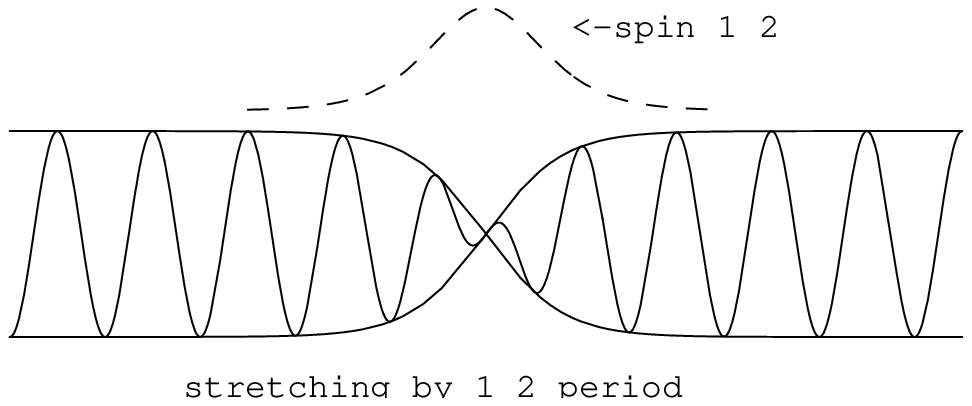}
\epsfxsize .7\hsize
\caption{Amplitude soliton in the IC CDW}
\end{minipage}\hfill 
\begin{minipage}[c]{.45\linewidth}
\includegraphics*[width=5.5cm]{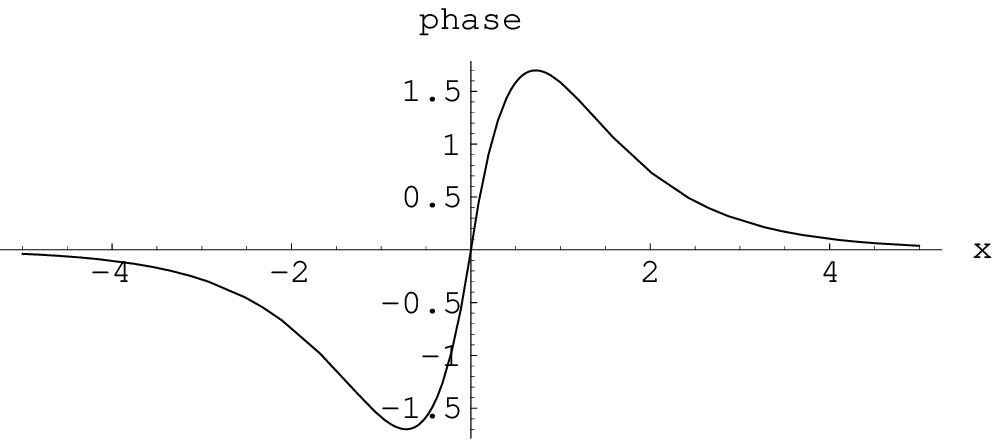}
\epsfxsize .35\hsize
\caption{Phase tails adapting the AS.}
\end{minipage}\hfill 
\end{figure}

\subsection {Spin-Gap cases: the quantum CDW and the superconductivity.}

We shall omit from consideration the case of repulsion which is relevant to the 
case of the incommensurate Spin Density Wave or to a hole within the 
antiferromagnetic  media which is important for doped Mott insulators. These 
cases were emphasized in previous publications \cite{braz-00}.
Here we shall concentrate upon systems with attraction which originate the gap 
and the discrete degeneracy in the spin channel. Firstly we shall generalize the 
description of the CDW solitons to the quantum model. Secondly we shall use the 
accumulated experience to arrive at our final goal: the spin carrier in the SC 
media.
 
$1D$ electronic systems are efficiently treated within the boson representation, 
see \cite{emery} for a review.  The variables can be chosen as $\varphi$ which 
is the analog of the CDW phase and $\theta $ which is the angle of the $SU2$ 
spin rotation. These phases are normalized in such a way that their increments 
divided by $\pi$ count the electronic charge and spin.

For the incommensurate electronic systems the Lagrangian can be written as 
\[
{\cal L}_{atr}\sim \{C_1(\partial \theta )^{2}+V\cos(2\theta )\}+C_2(\partial \varphi
)^{2} \, ; \,\, C_1,C_2=cnst 
\]
where $V$ is the backward exchange scattering and 
$(\partial f)^{2}=v^{-2}(\partial_t f)^{2}-(\partial_x f)^{2}$, 
$v\sim v_F$ is the velocity. 
Elementary excitations in $1D$ are the \textit{spinon} as a soliton 
$\theta=0 \rightarrow \theta =\pm\pi $, hence carrying the spin $\pm 1/2$,  and the 
gapless charge sound in $\varphi $. It is important to recall the alternative 
description in terms of conjugated phases. We shall need only the one for the 
charge channel which is the standard gauge phase $\chi$ of the 
superconductivity. Phases $\varphi$ and  $\chi$ are related (Efetov and Larkin 79) since 
their derivatives determine the same quantity - the current $j$: 
$\partial_t \varphi/\pi=j\sim\partial_x\chi $. The term 
$C_2(\partial \varphi)^{2}$ is duel to $\tilde C_2(\partial \chi)^{2}$
with $\tilde C_2\sim 1/C_2$

\subsubsection {The Quantum CDW.}

The CDW order parameter is 
$
O_{cdw}\sim \exp [i(Qx+\varphi )]\cos \theta 
$.
The spin operator $\cos \theta $ stands for what was the amplitude in the quasi-
classical description and at presence of the spinon it changes the sign as it 
was for $\Delta$.  
Hence for the CDW ordered state in a quasi $1D$ system the allowed configuration 
must be composed with two components: 
the spin soliton $\theta \rightarrow \theta +\pi $ 
and the phase wings  $\varphi \rightarrow \varphi +\pi $ 
where the charge $e=1$ is concentrated. Beyond the low dimensionality, a general 
view is:  \emph{the spinon as a soliton bound to the semi-integer dislocation 
loop}. 

\begin{figure}[htbf]
\includegraphics*[width=5.5cm]{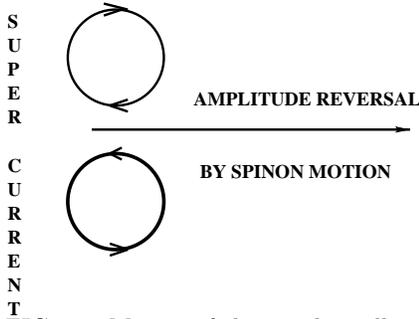}
\epsfxsize .85\hsize
\caption{
Motion of the topologically combined excitation in a spin-gap media. The string 
of the amplitude reversal of the order parameter created by the spinon is cured 
by the semi- vortex pair (the loop in $3D$) of the phase circulation. 
For the CDW case the curls are displacements contours for the half integer 
dislocation pair.
For the superconductivity the curls are lines of electric currents circulating 
through the normal core carrying the unpaired spin. 
}
\end{figure}

\subsubsection {The Singlet Superconductivity.}
For the Singlet Superconductivity the order parameter is 
$O_{sc}\sim \exp [i{\chi}]\cos \theta $.
In $D>1$ the elementary spin excitation is
composed with the \emph{spin soliton} $\theta \rightarrow \theta +\pi$
supplied with \emph{current wings} 
$\chi \rightarrow \chi +\pi $. 
The quasi $1D$ interpretation is that the \emph{spinon works as a
Josephson }$\pi -$\emph{\ junction} in the superconducting wire. The $\ 2D$
view is a pair of superconducting $\pi -$ 
{\it vortices sharing the common core where the unpaired spin is localized}. 
The $3D$ view is a 
{\it half flux vortex loop which collapse is prevented by the spin confined in its center}.

The solitonic nature of the spinon in the quasi $1D$ picture corresponds to the 
string of reversed sign of the order parameter left behind in the course of  the 
spinon motion. The spin soliton becomes an elementary fragment of the stripe 
pattern near the pair breaking limit (FFLO phase).
In a wire the $\pi$ wings of the spinon motion become the persistent current \cite{yak}. 

For this combined particle the  electronic quantum numbers are reconfined 
(while with different scales of localization).
But integrally over the cross-section the local electric current induced by the 
spinon  is compensated exactly by the back-flows at distant chains. 
This is a general property of the vortex dipole configuration constructed above.
Finally, the soliton as a state of the coherent media will not carry a
current and itself will not be driven by a homogeneous electric field. 

\section{Conclusions.}

Our conclusions have been derived for weakly interacting SC or CDW chains. 
Since the results are symmetrically and topologically characterized, they can be extrapolated 
to isotropic systems with strong coupling where a clear microscopical derivation 
is not available. Here the hypothesis is that instead of normal carries excited 
or injected above the gap, the lowest states are the symmetry broken ones 
described above as semiroton - spinon complex. This construction can be 
processed from another side considering a vortex configuration bound to an 
unpaired electron. Without assistance of the quasi one-dimensionality, 
a short coherence length is required to leave only a small number of intragap  levels in the vortex core. 

The existence of complex spin excitations in superconductors is ultimately related to robustness of the FFLO phase at finite spin polarization. It must withstand a fragmentation due to quantum or thermal melting at small spin polarization. Then any termination point of one stripe within the regular pattern (the dislocation) will be accompanied by the phase semiroton in accordance with the quasi $1D$ picture.

\end{document}